\documentclass{emulateapj}

\newcommand{\lya}{Ly$\alpha$}
\newcommand{\lyb}{Ly$\beta$}

\newcommand{\hi}{H~{\sc i}}
\newcommand{\oi}{O~{\sc i}}
\newcommand{\cii}{C~{\sc ii}}
\newcommand{\ciii}{C~{\sc iii}}
\newcommand{\civ}{C~{\sc iv}}
\newcommand{\siii}{Si~{\sc ii}}
\newcommand{\siiv}{Si~{\sc iv}}

\newcommand{\mgii}{Mg~{\sc ii}}
\newcommand{\feii}{Fe~{\sc ii}}

\newcommand{\heii}{He~{\sc ii}}
\newcommand{\mhi}{{\rm H \; \mbox{\tiny I}}}

\newcommand{\mciv}{{\rm C \; \mbox{\tiny IV}}}

\newcommand{\kms}{km~s$^{-1}$}

\newcommand{\mOm}{\Omega_{\rm m}}
\newcommand{\mOl}{\Omega_{\Lambda}}

\newcommand{\Ociv}{$\Omega_{\mciv}$}
\newcommand{\mOciv}{\Omega_{\mciv}}
\newcommand{\Nciv}{$N_{\mciv}$}

\newcommand{\fN}{$f(N_{\mciv})$}

\shorttitle{\civ\ at $z > 5.3$} 

\shortauthors{Becker et al.}

\begin{document}

\title{High Redshift Metals I.: The Decline of \civ\ at $z >
  5.3$\altaffilmark{1}}

\author{George D. Becker\altaffilmark{2,3}, Michael
        Rauch\altaffilmark{3}, Wallace L. W. Sargent\altaffilmark{4}}

\altaffiltext{1}{The observations were made at the W.M. Keck
  Observatory which is operated as a scientific partnership between
  the California Institute of Technology and the University of
  California; it was made possible by the generous support of the
  W.M. Keck Foundation.}
\altaffiltext{2}{Fellow, Kavli Institute for Cosmology, Madingley
  Road, Cambridge, CB3 0HA, UK; gdb@ast.cam.ac.uk}
\altaffiltext{3}{Carnegie Observatories, 813 Santa Barbara Street,
  Pasadena, CA 91101, USA; mr@ociw.edu}
\altaffiltext{4}{Palomar Observatory, California Institute of
  Technology, Pasadena, CA 91125, USA; 
  wws@astro.caltech.edu}

\begin{abstract}

We present the results from our search for \civ\ in the intergalactic
medium at redshifts $z = 5.3 - 6.0$.  We have observed four $z \sim 6$
QSOs with Keck/NIRSPEC in echelle mode.  The data are the most
sensitive yet taken to search for \civ\ at these redshifts, being 50\%
complete at column densities $\log{N_{\mciv}} \approx 13.4$.  We find
no \civ\ systems in any of the four sightlines.  Taking into account
our completeness, this translates into a decline in the number density
of \civ\ absorbers in the range $13.0 < \log{N_{\mciv}} < 15.0$ of at
least a factor 4.4 (95\% confidence) from $z \sim 2-4.5$, where the
number density is relatively constant.  We use our lack of detections
to set limits on the slope and normalization of the column density
distribution at $z = 5.3-6.0$.  The rapid evolution of \civ\ at these
redshifts suggests that the decrease in the number density may largely
be due to ionization effects, in which case many of the metals in the
$z \sim 4.5$ IGM could already be in place at $z \sim 5.3$, but in a
lower ionization state.  The lack of weak systems in our data,
combined with the presence of strong \civ\ absorbers along at least
one other sightline, further suggests that there may be large-scale
variations in the enrichment and/or ionization state of the $z \sim 6$
IGM, or that \civ\ absorbers at these redshifts are associated with
rare, UV-bright star-forming galaxies.

\end{abstract}


\section{Introduction}\label{sec:civ_intro}

Metal absorption lines are one of our most versatile tools for
studying the high-redshift Universe.  Just as the ionization history
of the IGM reflects the output of ionizing photons from star-forming
galaxies and quasars, the metal content of the IGM can be used to
infer the integrated and instantaneous rates of global star formation.
At very high redshifts, where the direct detection of galaxies becomes
increasingly challenging, metal absorption lines can serve as sign
posts to the ``typical'' galaxies that may be responsible for
reionizing the Universe at $z > 6$.  Perhaps most intriguingly, metal
lines can also serve as a probe of the reionization process itself.

A considerable amount of work on the high-redshift Universe has
focused on cosmic reionization, yet the timing and mechanisms by which
the IGM becomes permanently ionized remain poorly understood.  The
three main observational constraints on reionization are (1) the
electron optical depth measured from the polarization of the cosmic
microwave background (CMB), (2) the evolution of the high-redshift
\lya\ forest, and (3) the luminosity functions of high-redshift
\lya-emitting galaxies (LAEs).  Five-year {\it WMAP} measurements of
the CMB are consistent with an instantaneous reionization at $z_r \sim
11$, but are also compatible with an extended reionization process
\citep{dunkley08}.  The rapid evolution of the mean transmitted
\lya\ flux in the spectra of $z \gtrsim 5.7$ quasars has been used to
infer that reionization may have ended as recently as $z \sim 6$
\citep{becker01,white03,fan02,fan06,gnedin06}.  The quasar results
remain controversial, however
\citep[e.g.,][]{songaila04,ohfur05,lidz06,becker07}.  \lya\ saturates
for even a tiny neutral fraction \citep[$f^{V}_{\mhi} \sim
  10^{-3}$;][]{fan02}, making it extremely challenging to measure the
ionization state of the IGM at these redshifts directly from the
\lya\ forest.  Finally, the lack of a sharp decline in the luminosity
function of LAEs from $z \approx 5.7$ to $z \approx 6.5$ suggests that
most of the volume of the IGM is already highly ionized by $z \sim
6.5$.  \citep[e.g.,][but see Kashikawa et
  al.~2006]{mr04,mr06,hu06,dijkstra07}.  Interpreting the LAE results
is somewhat complicated, however, as \lya\ photons may escape through
locally ionized bubbles
\citep[e.g.,][]{haiman05,wyithe05,furlanetto06} or by redshifting via
galactic winds before passing through the IGM \citep{santos04}.

Quasar metal absorption lines provide a complimentary tool for
studying the ionization state of the IGM.  In a reionization scenario
where galaxies provide most of the ionizing flux, overdense regions of
the IGM should be the first to become enriched, yet may remain
significantly neutral until the end of reionization due to the short
recombination times at high densities.  ``Forests'' of low-ionization
absorption lines such as \oi~$\lambda 1302$, \siii~$\lambda 1260$, and
\cii~$\lambda 1334$ may therefore be present prior to the completion
of reionization \citep{oh02}.  Our recent search for \oi\ using HIRES
spectra of $z > 5$ QSOs yielded mixed results \citep{becker06}.  While
an overdensity of low-ionization absorbers at $z \sim 6$ was found
along at least one sightline, SDSS~J1148$+$5251 ($z_{\rm Q} = 6.42$),
no enhancement was seen towards SDSS~J1030$+$0524 ($z_{\rm Q} =
6.30$), which has a complete Gunn-Peterson trough spanning $\sim
80$~comoving Mpc \citep{white03}.  These differences suggest there may
be large-scale variations in the enrichment and/or ionization state of
the IGM at $z \sim 6$, and perhaps a strong redshift evolution in the
ionization state of metal absorption systems.  The metal-enriched,
low-ionization regions towards SDSS~J1148$+$5251 may represent the
final stages of reionization.  Alternatively, they may arise by chance
from multiple pockets of enriched gas near galaxies aligned along a
filament.

In order to effectively use metal lines as probes of reionization, the
roles of enrichment and ionization must be separately understood.  A
quasar sightline that passes through a largely neutral IGM may not
show an \oi\ forest if the IGM has not yet been enriched.  At lower
redshifts, highly-ionized species such as \civ\ and \siiv\ have been
the primary tracers of IGM metallicity \citep[e.g.,][]{meyer87, ssb88,
  petitjean94, cowie95, ellison99, schaye03, pettini03, aguirre04,
  simcoe04, scannapieco06}.  Both the comoving mass density of and the
column density distribution of \civ\ remain roughly constant over at
least $1.5 \lesssim z \lesssim 4.5$ \citep{songaila01}.  Since the
fraction of carbon in the form of \civ\ varies depending on the
ionization state of the gas, however, a constant \civ\ density does
not necessarily mean that the total IGM metallicity remains unchanged
\citep[e.g.,][]{oppenheimer06}.  Nevertheless, some fraction of IGM
metals are clearly in place by $z \sim 4.5$.

The current constraints on the abundance of metals in the IGM at $z >
5$ are mixed.  \citet{songaila05} found tentative evidence for a
decrease in the comoving mass density of \civ\ from $z \sim 4$ to 5.
To measure the metal content of the IGM at even redshifts the
\civ~$\lambda1548,1551$ doublet must be pursued into the
near-infrared.  Preliminary searches by \citet{simcoe06} and
\citet{r-w06} found evidence for a comoving \civ\ mass density at $z
\sim 6$ that is consistent with measurements at $z \sim 2-4.5$.  In
addition, \citet{simcoe06} found tentative evidence that the
\civ\ column density distribution may also remain constant out to $z
\sim 6$.  These works examined only two sightlines, however, and were
generally sensitive only to the rare, high-column density
($\log{N_{\mciv}} \gtrsim 13.8$) systems that dominate the comoving
mass density.  A complete census of \civ\ at $z \sim 6$ requires a
broader, more sensitive search.

In this series of papers, we present the results from a two-part
search for intergalactic metals at $z \sim 6$.  In this paper, we
describe our survey for \civ\ performed with Keck/NIRSPEC in
high-resolution mode.  This is the most sensitive search for \civ\ at
$z > 5.3$ yet performed, partly due to the high resolution ($R \approx
13,000$), which is better suited to search for narrow \citep[$b \sim
  10$~\kms,][]{rauch96} lines than previous studies.  By covering
multiple sightlines with high sensitivity, we are able to place strong
constraints on the abundance of the kinds of highly-ionized absorbers
that are common at lower redshifts.  In the following paper, we will
expand our search for low-ionization systems initially presented in
\citet{becker06}.  The combined results will provide unique constrains
not only on the ionization state of the IGM, but on the star formation
history of the Universe at $z > 5$.

The rest of the paper is organized as follows.  The data are presented
in \S2.  In \S3 we give the results of our search for \civ, along with
our completeness estimates.  We show that the number density of
\civ\ absorbers at $z > 5.3$ has declined sharply from $z \sim 2-4.5$
in \S4, and give constraints on the column density distribution and
the contribution of \civ\ to the closure density.  In \S5 we discuss
possible reasons for the decrease in \civ\ with redshift and the
significance of our lack of detections in light of the strong systems
previously reported along other sightlines.  Finally, our conclusions
are presented in \S6.  Throughout this paper, we assume $\mOm =
0.274$, $\mOl = 0.726$, and $H_0 = 70.5$~\kms\ \citep{komatsu08}.

\begin{deluxetable*}{lccccc}
   \tablecolumns{6}
   \tablecaption{Summary of NIRSPEC Echelle Observations} 
   \tablehead{\colhead{QSO\tablenotemark{a}} & 
              \colhead{redshift} &
              \colhead{$t_{\rm exp}$~(hrs)} & 
              \colhead{$S/N$\tablenotemark{b}} &
              \colhead{$\Delta z_{\mciv}$\tablenotemark{c}} &
              \colhead{$\Delta X_{\mciv}$\tablenotemark{d}} } 
   \startdata
      SDSS~J0002$+$2550 & 5.82 & 12.0 & $11.4-16.7$ & 0.458 & 2.33 \\
      SDSS~J0818$+$1722 & 6.00 & 11.0 & $10.3-14.2$ & 0.638 & 3.29 \\
      SDSS~J0836$+$0054 & 5.80 & 11.5 & $14.5-21.5$ & 0.438 & 2.23 \\
      SDSS~J1148$+$5251 & 6.42 & 12.0 & $10.9-15.6$ & 0.679 & 3.44
   \enddata 
   \tablenotetext{a}{QSO names given as SDSS Jhhmm$+$ddmm}
   \tablenotetext{b}{Middle 50\% range of signal-to-noise per
     resolution element values for pixels within the wavelength
     interval used to search for \civ}
   \tablenotetext{c}{Total redshift interval used to search for \civ}
   \tablenotetext{d}{Total absorption path length interval used to
     search for \civ \vspace{0.05in}}
   \label{tab:ne_obs}
\end{deluxetable*}

\section{The Data}\label{sec:civ_data}

\subsection{NIRSPEC Echelle spectra}

We obtained high-resolution, near-IR spectra of four $z \sim 6$ QSOs
with Keck/NIRSPEC \citep{mclean98} between February 2007 and February
2008.  The spectrograph was used in echelle mode in the NIRSPEC-1 (Y)
band, which provides continuous wavelength coverage over $\lambda
\approx 0.95 - 1.13$~\micron\ in a single setting.  Our observations
are summarized in Table~\ref{tab:ne_obs}.  Individual exposures were
1800 seconds, with total integration times of 11-12 hours per object.

Data reduction was performed using a custom set of IDL routines
tailored to the technical challenges of observing faint targets in the
infrared.  The NIRSPEC detector, a 1024x1024 ALLADIN-3 InSb array, has
good sensitivity in the near infrared but also a high dark current.
The sky emission lines are widely separated in echelle mode, leaving
the dark current as the dominant source of noise provided that other
sources such as variable bias levels, hot pixels, and numerous
cosmetic blemishes can be properly removed.

Rather than nod-subtract our exposures, which automatically multiplies
the noise from the sky and the dark current by a factor of root two,
we compiled a low noise dark frame by taking a large set of dark
exposures with the same exposure time as our science frames.  This was
used to remove dark current features and other blemishes prior to sky
subtraction.  The exposures were flat fielded using internal flats.
In order to correct for pixel-to-pixel variations separately from the
high-frequency fringing introduced by interference between the order
sorting and blocking filters, the fringe pattern was modeled and
removed from the two dimensional flats.  Each science exposure was
then divided by the ``pixel'' flat, as well as by the fringe model
after it was scaled in intensity and shifted on the detector to match
the exposure.

The sky was modeled using a b-spline fit and removed from each
exposure using optimal sky subtraction techniques \citep{kelson03}.
The residual bias level from each of the thirty-two independent
readout channels was modeled and removed both at the start of the
reduction and again after a first-pass sky subtraction.  Similar care
was taken to remove the residual bias in the dark frames.

In order to maximize the signal-to-noise ratios of the final spectra,
we performed simultaneous multiple spectral extraction on all
exposures for a given object.  The object spatial profile was modeled
independently for each exposure.  A response function derived from a
standard star exposure was also applied to the counts in the 2-D
frames.  A single 1-D spectrum was then extracted from all exposures
of a given object simultaneously using optimal weighting
\citep{horne86}.  This technique allowed us to best reject spurious
pixels while preserving as much information as possible from the 2-D
arrays, which were individually very noisy even after careful
reduction.  The 1-D spectra were then normalized using a
slowly-varying spline fit to the continuum.  Telluric absorption
correction was done using a model of the atmospheric transmission in
the Y band \citep{hinkle03}.  This produced better results than using
our standard star spectra, which contained some residual
high-frequency ringing as a result of the fringing.  Even so, we have
restricted our analysis to the center of the transmission window
($9850 - 10920$~\AA), where there is very little atmospheric
absorption.  Finally, we scaled the formal error arrays so that the
distribution of deviations of the flux from the continuum fit,
normalized by the error array, was well fit by a Gaussian with unit
variance for each object.  The typical resulting increase in the error
array was $\sim 30\%$, which is a reasonable estimate for the amount
of noise that could not be well characterized in the 2-D frames.

\begin{figure*}
   \epsscale{1.15} 
   \centering 
   \plotone{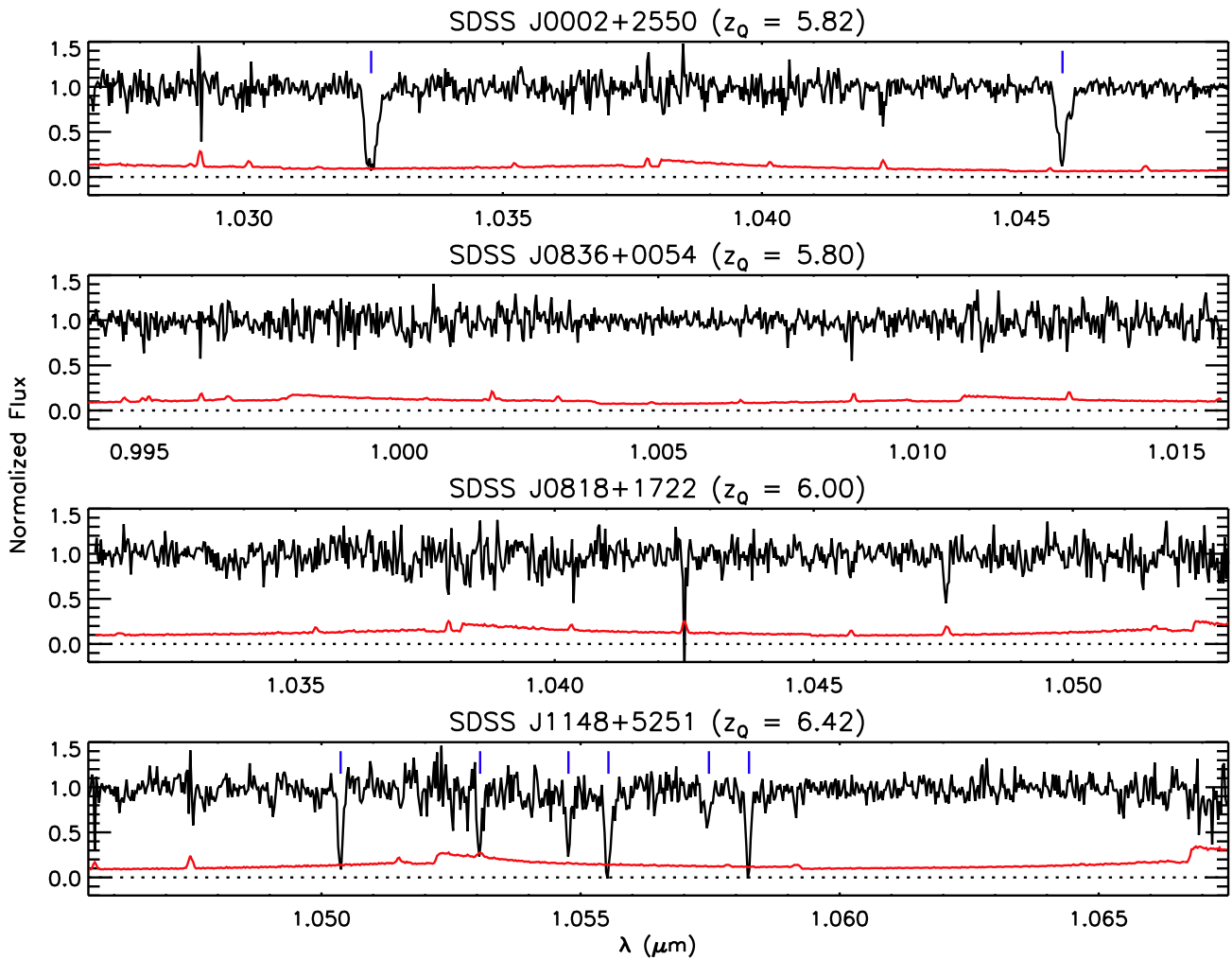}
   \caption{Example sections of the final 1-D normalized NIRSPEC
     echelle spectra.  The formal error arrays are shown in red.
     Genuine absorption lines are identified with tick marks.  The
     lines in the spectrum of SDSS~J0002$+$2550 are
     \feii~$\lambda2344,2374$ at $z = 3.404$.  Those in the spectrum
     of SDSS~J1148$+$5251 are three sets of \mgii~$\lambda2796, 2804$
     doublets at $z \approx 2.77$.  Features in the flux that coincide
     with spikes in the error arrays are typically residuals from the
     subtraction of strong skylines. \label{fig:example_sections}}
\end{figure*}

Representative sections of the final spectra are shown in
Figure~\ref{fig:example_sections}.  The data tend to become noisy at
the ends of the echelle orders, which are marked by a step in the
error array.  The measured resolution is $R \approx 13000$ ($FWHM
\approx 23$~\kms).  Narrow metal lines can clearly be seen, though
these are at much lower redshifts.  Significantly, the noise
properties of the data are well-behaved, with a scatter as a function
of wavelength that is well-described by the scaled error array.  This
is essential for correctly estimating our sensitivity.

\subsection{Optical spectra}

We have also acquired high-resolution optical spectra of $z \sim 6$
QSOs.  Some of these data were presented in \citet{becker06,becker07},
and the new observations will be discussed more thoroughly in the next
paper.  Here we will use Keck HIRES \citep{vogt94} spectra of
SDSS~J0836$+$0524 and SDSS~J1030$+$0524 ($z_{\rm Q} = 6.30$) to help
evaluate candidate \civ\ detections along those sightlines.  The data
were reduced using a custom IDL pipeline featuring optimal sky
subtraction.  As with the NIRSPEC data, simultaneous optimal
extraction was also used to maximize the signal-to-noise ratio of the
final spectra.  Prior to extraction, the 2-D counts were flux
calibrated using a response function generated from a standard star.
Telluric absorption correction using a standard star was also applied
to the 2D frames.  Velocity offsets between the standard and the
science frames, as well as variations in the strength of the ${\rm
  H_2O}$ and ${\rm O_2}$ absorption bands were determined from
preliminary extractions of groups of individual exposures taken near
the same time.  Since SDSS~J1030$+$0524 was observed on multiple runs
at different times of the year, shifts in the relative velocity of the
object allowed residuals from atmospheric absorption and emission to
be especially well rejected.

\section{Results of \civ\ Search}

We searched for \civ\ doublets in our NIRSPEC data both by eye and
using an automated technique described below.  Although the spectra
were corrected for telluric absorption, we limited our search to the
regions between 9820~\AA\ and 10920~\AA\ ($5.34 \le z_{\mciv} \le
6.04$), which is largely free of atmospheric absorption.  The total
redshift interval along each sightline is given in
Table~\ref{tab:ne_obs}.  In cases where the QSO \civ\ emission line
fell in the bandpass, we included the entire region blueward of the
line center.  While it is customary to exclude a narrow region near
the QSO redshift in order to avoid proximity effects, this region is
also typically enhanced with \civ\ absorbers
\citep[e.g.,][]{vestergaard03,nestor08}.  Our conclusions below are
therefore most likely enhanced by including it in our analysis.  In
any case, the proximity region comprises only a small part ($\sim
10$\%) of the total path length towards two of our four QSOs.

\begin{figure}
   \epsscale{1.0} 
   \centering 
   \plotone{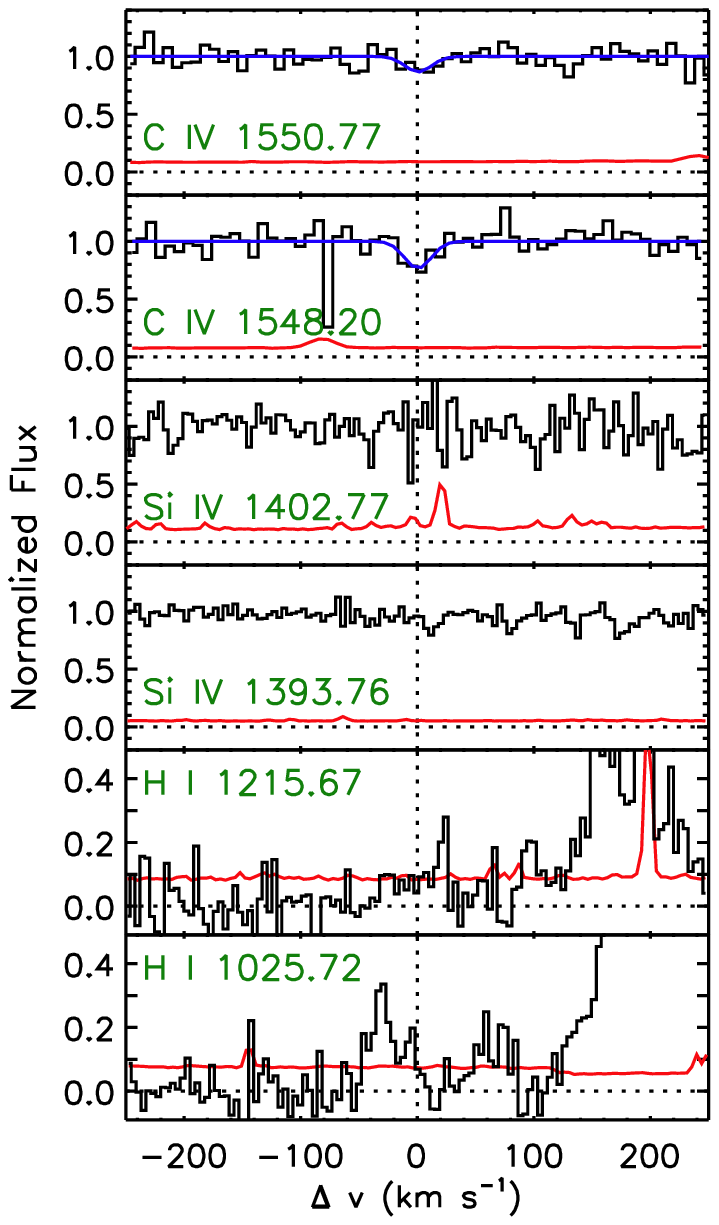}
   \caption{Candidate \civ\ system at $z=5.409$ towards
     SDSS~J0836$+$0054.  Histograms show the normalized flux, which is
     from NIRSPEC for \civ\ and HIRES (binned to 4.2~\kms\ for
     display) for \siiv\ and \hi.  Red lines are the formal error
     arrays.  The blue line is a Voigt profile fit to the candidate
     \civ\ absorption, with $N_{\mciv} = 10^{13.06}~{\rm cm^{-2}}$.
     The presence of \hi\ \lya\ and \lyb\ transmission at the same
     redshift strongly indicates that this is a false
     detection.\label{fig:civ_z5.409}}
\end{figure}

Both the manual search and the follow-up automated search revealed
only a single candidate \civ\ absorber, which is at $z=5.4091$ towards
SDSS~J0836$+$0054.  This system is plotted in
Figure~\ref{fig:civ_z5.409}.  Voigt profile fitting produced a
best-fit column density $\log{N} = 13.06 \pm 0.12~{\rm (cm^{-2})}$ and
a Doppler parameter $b = 5.8 \pm 6.3$~\kms (i.e., unresolved).  Using
the HIRES spectrum of SDSS~J0836$+$0054, we can measure \siiv\ and
\hi\ \lya\ and \lyb\ at the same redshift.  There is no detectable
\siiv, although this does not necessarily rule out a legitimate
\civ\ absorber.  More significantly, the HIRES data show transmission
in \lya\ and \lyb\ at this redshift.  Since \civ\ absorbers of this
column density are almost always accompanied by saturated
\lya\ absorption, at least at lower redshifts \citep[e.g.,][but see
  Schaye et al.~2007]{ellison99,simcoe04}, we can safely conclude that
this system is not real.  Thus, both our manual and automated searches
produced no \civ\ detections.

\subsection{Completeness}

To assess the significance of our lack of \civ\ detections we have
estimated our completeness as a function of column density using a
Monte Carlo method.  Artificial \civ\ doublets were inserted randomly
into the data and an automated algorithm was used to determine whether
these systems could be detected.  We generated $10^{4}$ doublets for
each sightline, ranging in column density from $10^{12}$ to
$10^{15}~{\rm cm^{-2}}$.  Doppler parameters were selected from the
measured distribution of \citet{rauch96}.

In order to be considered a detection, we first required the flux
decrement to exceed the local error at the wavelength of both lines of
the doublet after the spectra were smoothed by the instrumental
resolution (${\rm FWHM} = 23$~\kms).  We then fit independent
Gaussians to each line (using the unsmoothed data), and evaluated the
reliability of the fits and the agreement between the two lines.  Our
tolerances were stricter for the stronger $\lambda 1548$ transition.
For a satisfactory fit, the absolute velocity uncertainty was required
to be less than 0.5 (1.0) times the resolution for the $\lambda 1548$
($\lambda 1551$) transition.  We further required that the fitted flux
decrement be greater than 3.0 (2.0) times the fit uncertainty, and the
line width be greater than both 3.0 (1.0) times the fit uncertainty
and 0.5 times the resolution.  A ``detection'' was only counted when
the line fits were also consistent with being a doublet.  In order to
exclude potential false positives, we set stricter requirements for
weaker lines, since strong lines are typically easy to identify by
eye.  When the peak optical depth of the $\lambda 1548$ transition was
less (greater) than 1.0, we required that the velocity offset between
the two fits be less than both 1.0 (2.0) times the resolution and 2.0
(3.0) time the combined uncertainty in the fitted velocities.  The
line widths were required to agree to within half the width of the
narrower line, or three times the combined uncertainty.  In all cases,
however, the line widths could not disagree by more than a factor of
two.  Finally, if the peak optical depth of one or both lines was less
than one, we required the ratio of the optical depths $\tau^{\rm
  peak}_{\lambda1548} / \tau^{\rm peak}_{\lambda1551}$ to be between
0.8 and 3.0.

Defining criteria for an automated search is a somewhat subjective
process, as selection thresholds and goodness-of-fit requirements will
vary depending on the line fitting method.  The above procedure was
designed to give the same result as a visual evaluation of the
synthetic lines as often as possible.  In general, this meant applying
greater scrutiny to weaker systems in order to avoid false positives.
We found very good agreement overall between the fraction of systems
identified by eye and by the automated method.

\begin{figure*}
   \epsscale{0.8} 
   \centering 
   \plotone{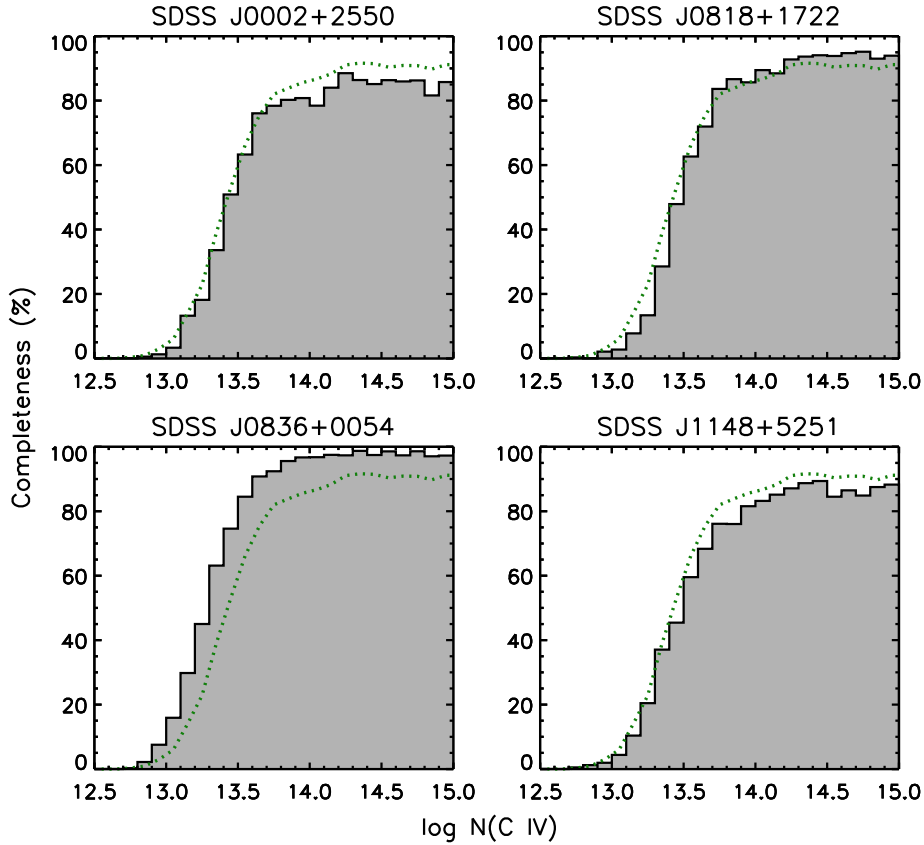}
   \caption{Percentage of recovered artificial \civ\ doublets as a
     function of column density.  Histograms give the completeness for
     individual sightlines.  The dotted line is the
     path length-weighted mean for the entire sample.  The completeness
     does not go to 100\%, even for large column densities, because of
     the non-negligible chance that of one of the \civ\ components may
     coincide with a lower-redshift absorption line, a residual from a
     subtracted skyline, or some other unusable part of the
     spectrum. \label{fig:civ_completeness}}
\end{figure*}

\begin{figure}
   \epsscale{1.0} 
   \centering 
   \plotone{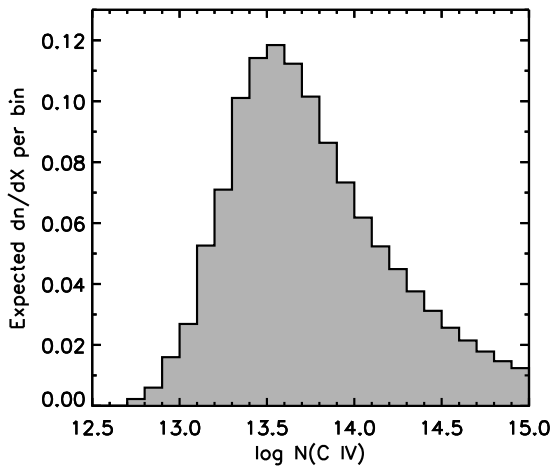}
   \caption{Expected \civ\ detection rate for a column density
     distribution that is unchanged from $z \sim 2-4.5$.  We computed
     the expected number of \civ\ systems per unit redshift path
     interval per log column density bin by integrating \fN\ from
     \citet{songaila01} against our path length-weighted mean
     completeness (Figure~\ref{fig:civ_completeness}).  Over our
     entire sample, we would expect to detect 13.2 systems with $13.0
     \le \log{N_{\mciv}} \le 15.0$. \label{fig:civ_expected}}
\end{figure}

Our completeness results are shown in
Figure~\ref{fig:civ_completeness}.  We are most complete at
$\log{N_{\mciv}} > 13.7$, though we are still 50\% complete in the
overall sample at $\log{N_{\mciv}} \approx 13.4$, and even retain some
sensitivity at $\log{N_{\mciv}} = 13.0$.  As discussed below, the fact
that we are significantly complete over a wide range of column
densities allows us to put strong constraints on the column density
distribution of \civ\ absorbers.

We also evaluated the number of false positive detections we should
expect.  Using the same automated search method, we searched the data
for ``fake'' doublets with velocity separations that were offset from
the \civ\ doublet separation.  In each trial we adjusted the doublet
separation by twice the instrumental resolution, omitting separations
that were similar to true doublets such as \mgii~$\lambda2796,2804$,
or other plausible combinations of intervening absorption lines.  In
twenty trials we detected seven false doublets, which gives an
expectation value of 0.35 for the entire data set.  This is consistent
with the single false \civ\ detection described above, although in
that case we were able to rule it out based on the presence of
transmitted flux in the \lya\ forest.

\section{Constraints on \civ\ in the $z \sim 6$ IGM}

\subsection{Change in the number density of absorbers from $z \sim 3$}

We can use our completeness estimates to calculate the number of
detections we should expect if the column density distribution has not
evolved from lower redshifts.  It is conventional to fit a power law
of the form
\begin{equation}
   \label{eq:f}
   f(N_{\mciv}) \equiv \frac{\partial ^2{\mathcal N}}{\partial N_{\mciv}\,
     \partial X } = B \left( \frac{N_{\mciv}}{N_{0}}
   \right)^{-\alpha}.
\end{equation}
Here, $X$ is the absorption path length interval, defined such that,
for a non-evolving population of sources, $d{\mathcal N}/dX$ remains
constant with redshift.  For $\Omega = 1$,
\begin{equation}
\frac{dX}{dz} = \frac{(1+z)^2}{\sqrt{\mOm(1+z)^3 + \mOl}} \, .
\end{equation}
Fitting a large sample of absorbers at $z = 2.90-3.54$,
\citet{songaila01} found a slope $\alpha = 1.8 \pm 0.1$ for systems
with $\log{N_{\mciv}} > 13.0$, with a normalization $B = 10^{-12.67}$
(corrected for our adopted cosmology) at $N_{0} = 10^{13.0}~{\rm
  cm^{-2}}$.  Indeed, they found that this distribution is nearly
invariant out to at least $z \approx 4.5$.  \citet{songaila05} found a
slightly shallower slope ($\alpha = 1.7$) and a higher normalization
when fitting the distribution of ``pseudoclouds'' identified by an
automated method.  We will compare our results to the
\citet{songaila01} fit, however, since the search method is more
similar to the one used here.

Figure~\ref{fig:civ_expected} shows our mean completeness function
convolved with the column density distribution of \citet{songaila01}.
We would expect the largest number of detections near $\log{N_{\mciv}}
\sim 13.5$, but there are significant tails towards both ends of the
column density range.  In total, we would expect to detect 13.2
systems with $\log{N_{\mciv}}$ between 13.0 and 15.0 along all four
sightlines if \fN\ was unchanged from $z \sim 3$.  For zero
detections, the 95\% and 99\% 1-sided confidence limits on the mean
expected number are 3.0 and 4.6, respectively
\citep[e.g.,][]{gehrels86}.  This translates into a 95\% confidence
limit of at least a factor of 4.4 decline in the number density of
\civ\ systems from $z \sim 3$ to $z = 5.3 - 6.0$, or a 99\% limit of a
factor of 2.9.  We note that with the \citet{songaila05} \fN\ we would
have expected 21 systems.  Similarly, using \fN\ from a fit to the
\civ\ distribution in nine high-quality HIRES spectra from
\citet{boksenberg03}, we would have expected 17 lines.  Adopting the
\citet{songaila01} column density distribution may therefore be
somewhat conservative.

\subsection{Previously studied sightlines}

\begin{figure}
   \epsscale{1.00} 
   \centering 
   \plotone{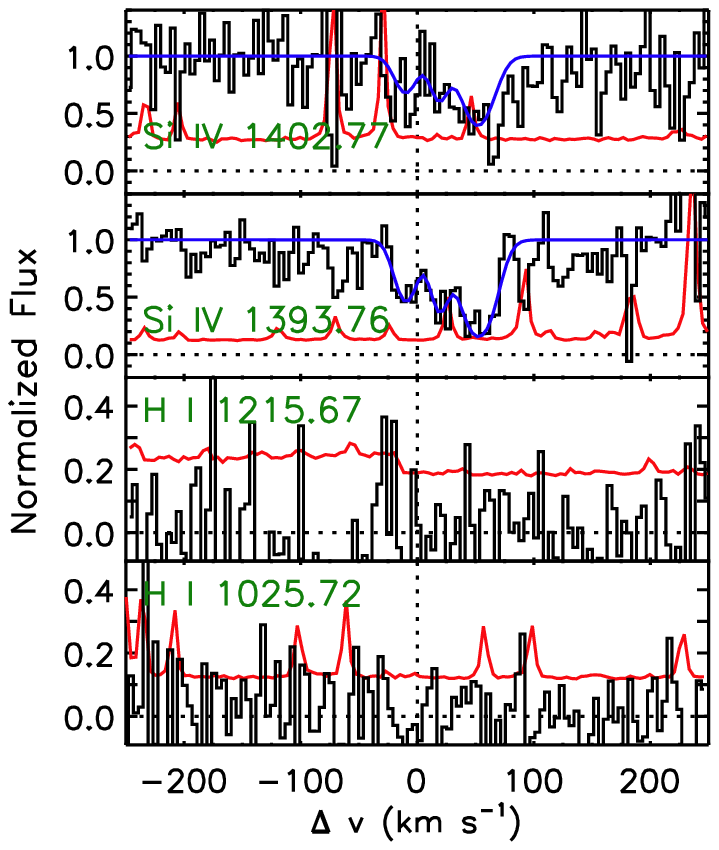}
   \caption{HIRES data showing the \siiv\ and \hi\ absorption profiles
     for the candidate \civ\ system at $z=5.724$ towards
     SDSS~J1030$+$0524 \citep{r-w06}.  The spectra have been binned
     using 4.2~\kms\ pixels for display.  The blue line is a Voigt
     profile fit to the \siiv\ using three components
     (Table~\ref{tab:siiv_z5.724}).  While the data are too noisy to
     determine the exact kinematic structure, the \siiv\ is clearly
     not saturated, and therefore the total fitted column density does
     not change significantly when more components are
     added.  \label{fig:siiv_z5.724}}
\end{figure}

Given the large drop in the number density of absorbers we observe, it
is worth reviewing the results from the initial searches for \civ\ at
$z > 5.3$.  \citet{simcoe06} obtained Gemini/GNIRS spectra of two $z
\sim 6$ QSOs, SDSS~J1306$+$0356 ($z_Q = 6.00$) and SDSS~J1030$+$0524
($z_Q = 6.27$).  \citet{r-w06} also observed SDSS~J1030+0524 with
VLT/ISAAC.  Simcoe found a number of low-column density absorbers
along both sightlines.  As the author points out, however, the
majority of these detections are near the detection limit and are
consistent with the expected number of false detections.  They found
at least one secure absorber, which occurs towards SDSS~J1030$+$0524
at $z = 5.829$ with $\log{N^{\rm total}_{\mciv}} = 13.8$.  Ryan-Weber
et al. also detected this system, albeit marginally (they measure a
much higher column density, but this may be due to blending of the
$\lambda 1551$ transition with an unrelated line that is visible in
the GNIRS spectrum).  They further detect a system at $z = 5.724$ that
is unclear in the GNIRS data.  We have checked the authenticity of the
$z = 5.724$ system by searching for related lines in our HIRES
spectrum of SDSS~J1030$+$0524.  The optical data, shown in
Figure~\ref{fig:siiv_z5.724}, show \siiv\ absorption at the redshift
of the candidate \civ.  We confirm the large ($\sim 60$~\kms) velocity
width measured by \citet{r-w06}.  At HIRES resolution, the
\siiv\ absorption is clearly asymmetric and best fit using multiple
components (see Table~2).  Assuming that the \civ\ absorption has the
same kinematic structure and relative component strength as the \siiv,
we were able to reproduce the Ryan-Weber et al. fit using a total
\civ\ column density very similar to their measured $\log{N_{\rm civ}}
= 14.4$.  Thus, there do not seem to be any hidden saturation effects
in the ISAAC data.

In summary, in the two sightlines studied by \citet{simcoe06} and
\citet{r-w06} there are two confirmed $z \sim 6$ \civ\ systems.  These
lie along a single sightline, and they are both strong
($\log{N_{\mciv}} > 13.8$) systems.  The weaker candidates reported by
\citet{simcoe06} are broadly consistent with their expected false
positive rate.

\begin{deluxetable}{lrr}
   \tabletypesize{}
   \tablewidth{0pt}
   \tablecaption{Voigt Profile Fit to \siiv\ at $z=5.724$ towards
                 SDSS~J1030$+$0524}
   \tablehead{ \colhead{$z$} & \colhead{$b$~(\kms)} & 
               \colhead{$\log{N}~{\rm (cm^{-2})}$} }
   \startdata
      $5.7235623 \pm 0.000049$  &  $11.8 \pm  3.6$  &  $12.97 \pm 0.10$ \\
      $5.7242061 \pm 0.000063$  &  $ 9.2 \pm  4.2$  &  $12.98 \pm 0.17$ \\
      $5.7249523 \pm 0.000042$  &  $16.2 \pm  2.7$  &  $13.47 \pm 0.06$
   \enddata
   \tablecomments{These are the best fitting parameters for a three
     component fit to \siiv\ in the HIRES data.  Increasing the number
     of components does not significantly change the total column
     density.}
   \label{tab:siiv_z5.724}
\end{deluxetable}

\subsection{Constraints on the column density distribution}

\begin{figure}
   \epsscale{1.1} 
   \centering 
   \plotone{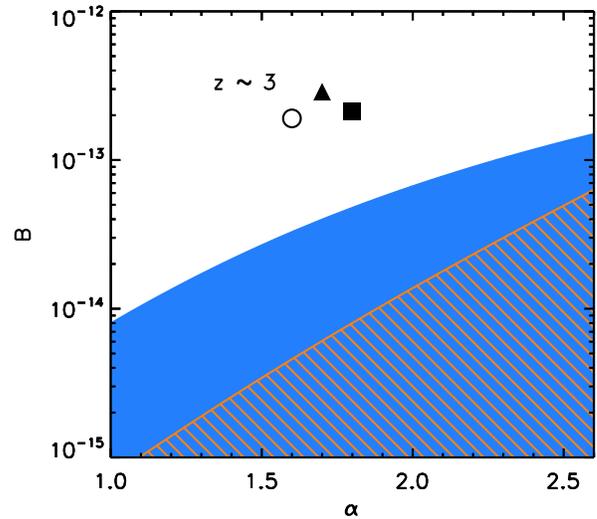}
   \caption{Limits on the column density distribution, $f(N_{\mciv}) =
     B N_{\mciv}^{-\alpha}$, of \civ\ absorbers at $z = 5.3 - 6.0$ in
     the range $13.0 < \log{N_{\mciv}} < 15.0$.  The solid shaded
     regions show the range of normalizations allowed at the 95\%
     confidence level for a given power-law slope based on the lack of
     detections in our NIRSPEC data.  Separately, the line-filled
     regions are excluded at the 95\% level based on the fact that
     there are at least two known systems with $\log{N_{\mciv}} \ge
     13.8$ along six lines of sight (see text).  Symbols show fitted
     parameters for \fN\ at $z \sim 3$, corrected for our adopted
     cosmology; {\it square}: \citet{songaila01}, {\it triangle}:
     \citet{songaila05}, {\it circle}:
     \citet{boksenberg03}.\label{fig:civ_plaw}}
\end{figure}

We can set joint constrains on the slope and normalization of a
power-law distribution of \civ\ column densities.  For a given slope
$\alpha$, we have calculated the 95\% limits on $B$ by adjusting the
normalization such that we would expect to detect 3.0 \civ\ systems
along our four sightlines.  In calculating $B$, we take into account
our measured completeness, and restrict ourselves to the number of
expected systems between $\log{N_{\mciv}} = 13.0$ and 15.0, for
reasons discussed below.  The results are shown in
Figure~\ref{fig:civ_plaw}.  As stated above, if $\alpha$ remains
unchanged from $z \sim 3$, then the normalization must decrease by at
least a factor of $\sim 4$.  Alternatively, \fN\ could steepen (larger
$\alpha$) while maintaining a somewhat higher normalization.  This
would effectively ``hide'' a greater fraction of the expected systems
below our detection limit.  It is also possible that \fN\ could become
more shallow, shifting the distribution towards stronger lines,
provided that the normalization decreases more sharply.  In
Figure~\ref{fig:civ_plaw} we have only shown the results for $\alpha
\ge 1$.  Formally, $\alpha$ could be less than one, but this would
imply more absorbers per decade in \Nciv\ with increasing \Nciv, which
is not observed for any other class of absorbers.  We further note
that low values of $\alpha$ strongly require the assumed high-end
cutoff in \Nciv\ at $10^{15.0}~{\rm cm^{-2}}$.

We can further use the detections of \citet{simcoe06} and
\citet{r-w06} to set lower limits on the normalization of \fN.  The
total path length surveyed by those authors and in this work is $\Delta
X = 17.4$.  For two detections, the 1-sided 95\% lower limit on the
expected mean number is 0.36.  In order to set a lower limit on $B$,
we assume 100\% completeness and adjust the normalization such that we
would expect at least this many systems with $\log{N_{\mciv}} =
13.8-15.0$ along all six sightlines.  The results are shown in
Figure~\ref{fig:civ_plaw}.  Including this constraint still leaves a
significant degeneracy between $\alpha$ and $B$, but it greatly
reduces the range of allowed normalizations for a given power-law
slope.

\subsection{Limits on \Ociv}

\begin{figure}
   \epsscale{1.1} 
   \centering 
   \plotone{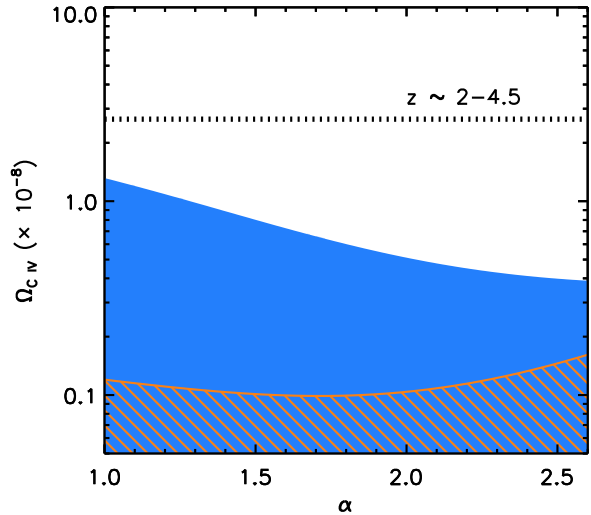}
   \caption{Limits the contribution of \civ\ to the closure density at
     $z = 5.3-6.0$, calculated by integrating over the allowed column
     density distributions in Figure~\ref{fig:civ_plaw}.  The solid
     shaded region shows the range of $\Omega_{\mciv}$ values allowed
     at the 95\% confidence level for a given slope $\alpha$, based on
     the lack of detections in our NIRSPEC data.  Separately, the
     line-filled regions are excluded at the 95\% level based on the
     fact that there are at least two known $\log{N_{\mciv}} \ge 13.8$
     systems along six lines of sight.  The values for
     $\Omega_{\mciv}$ were calculated by integrating
     Eq.~\ref{eq:omega} over $10^{13.0}~{\rm cm^{-2}} \le N_{\mciv}
     \le 10^{15.0}~{\rm cm^{-2}}$ (see text).  The horizontal dotted
     line shows $\Omega_{\mciv}$ at $z \sim 2-4.5$ calculated by
     integrating over the \citet{songaila01} \fN\ between the same
     bounds, which produces good agreement with published
     observations.  \label{fig:civ_omega_limits}}
\end{figure}

Our limits on the total \civ\ mass density are necessarily weaker than
our constraints on \fN.  For $\alpha < 2$, the integrated mass density
will be dominated by the highest column density absorbers, which may
be too rare to appear in our data.  Nevertheless, we can set limits on
\Ociv\ by integrating over the allowed column density distributions.
The contribution of \civ\ to the closure density $\rho_c$ can be
calculated as
\begin{equation}
   \label{eq:omega}
   \Omega_{\mciv} = \frac{m_{\mciv}}{\rho_c} \frac{H_0}{c} 
                 \int^{N_{\rm max}}_{N_{\rm min}}{
                  N_{\mciv} \, f(N_{\mciv}) \, dN_{\mciv}} \, ,
\end{equation}
where $m_{\mciv}$ is the mass of the \civ\ ion.  The choice of
integration limits is critical, but there are natural choices for
\civ.  At lower redshifts, \fN\ tends to become significantly
shallower at $\log{N_{\mciv}} \lesssim 13.0$
\citep[e.g.,][]{songaila05}.  Integrating the completeness-corrected
fits to \fN\ of \citet{songaila05} yields a contribution to
\Ociv\ below $10^{13.0}~{\rm cm^{-2}}$ that is $< 10$\% of the
contribution between $10^{13.0}$ and $10^{15.0}~{\rm cm^{-2}}$.  At
the other end, systems with $\log{N_{\mciv}} > 15.0$ are not typically
detected, even in large surveys \citep[e.g.,][but see Scannapieco et
  al.~2006]{songaila01,songaila05}.  There is likely to be another
break in the column density distribution above $\log{N_{\mciv}} \sim
15.0$, at least at moderate redshifts.  \citet{songaila01} would have
expected to detect $\sim 5$ systems with $15 < \log{N_{\mciv}} < 16$
between $z = 2.0$ and $z = 4.5$ if their power-law distribution held to
these column densities.  At $z \sim 2-4.5$, the measured $\mOciv \sim
3 \times 10^{-8}$ is well-reproduced by integrating over the
\citet{songaila01} \fN\ between $10^{13.0}$ and $10^{15.0}~{\rm
  cm^{-2}}$.  We therefore adopt $N_{\rm min} = 10^{13.0}~{\rm
  cm^{-2}}$ and $N_{\rm max} = 10^{15.0}~{\rm cm^{-2}}$ as our
integration limits.

\begin{figure}
   \epsscale{1.15} 
   \centering 
   \plotone{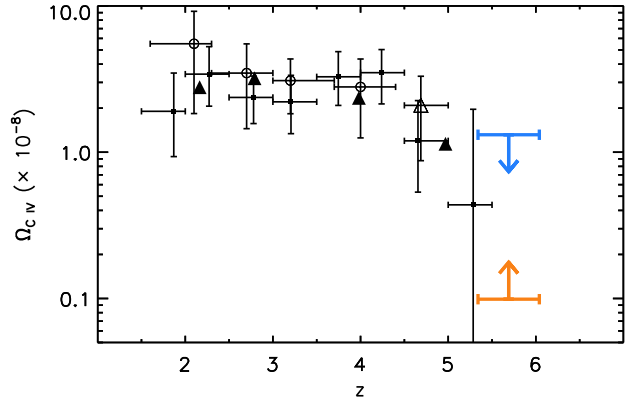}
   \caption{\Ociv\ as a function of redshift. The downward-pointing
     arrow shows the 1-sided 95\% upper limit on $\Omega_{\mciv}$ at
     $z = 5.3-6.0$ set by the lack of detections in our NIRSPEC data
     (see Figure~\ref{fig:civ_omega_limits}).  The limit assumes a
     power-law column density distribution, $f(N_{\mciv}) \propto
     N_{\mciv}^{-\alpha}$, with $\alpha \ge 1.0$ for $13.0 <
     \log{N_{\mciv}} < 15.0$.  Separately, the upward-pointing arrow
     shows the 95\% lower limit on \Ociv\ for $\alpha \ge 1.0$ set by
     the fact that there are at least two known $\log{N_{\mciv}} \ge
     13.8$ systems along six lines of sight.  Symbols are measurements
     from the literature, corrected for our adopted cosmology, with
     2-sided 90\% vertical error bars; {\it small squares}:
     \citet{songaila01}, {\it filled triangles}: \citet{songaila05},
           {\it open circles}: \citet{boksenberg03}, {\it open
             triangle}: \citet{pettini03}.\label{fig:civ_omega}}
\end{figure}

The results of integrating over \fN\ using the allowed values of the
power-law slope and normalization are shown in
Figure~\ref{fig:civ_omega_limits}.  The corresponding value at $z \sim
2-4.5$, obtained by integrating over \fN\ from \citet{songaila01}, is
shown as a horizontal line.  Our data at least demonstrate that
$\mOciv(z = 5.3- 6.0)$ must be $\lesssim 1.3 \times 10^{-8}$, or
$\lesssim 0.5 \; \mOciv(z \sim 2-4.5)$.  As expected, the largest
allowed values of \Ociv\ are for very shallow \fN, which permit there
to be rare, strong systems without also requiring a large number of
weaker systems.  If $\alpha$ remains unchanged from $z \sim 2-4.5$,
however, then \Ociv\ is limited to $\lesssim 6.0 \times 10^{-8}$, or a
factor of $\sim 4$ less than $\mOciv(z \sim 2-4.5)$.  Separately,
$\mOciv(z = 5.3-6.0)$ must be $\gtrsim 1 \times 10^{-9}$ in light of
the two detected systems towards SDSS~J1030$+$0524.  These limits are
compared to measurements from the literature in
Figure~\ref{fig:civ_omega}.  We note that our limits on \Ociv\ based
on the allowed shape of the column density distribution are consistent
with the results from a broader search for strong \civ\ systems along
a larger number of $z \sim 6$ sightlines (Ryan-Weber et al., in prep).

\section{Discussion}

Our primary goal in this paper has been to measure the evolution of
\civ\ absorbers at $z = 5.3-6.0$.  We have shown that the weak
($\log{N_{\mciv}} \gtrsim 13.0$) absorption systems that are
ubiquitous at $z < 4.5$ are less abundant by at least a factor of four
at $z > 5.3$.  At face value, this might suggest that the metal
content of the IGM is changing rapidly at these redshifts.  Several
factors, however, must be considered when translating the observed
incidence of \civ\ systems into a global IGM metallicity.  The number
density of highly ionized metal absorption systems at a given redshift
will depend not only on the mean metallicity of the IGM, but also on
the the spatial distribution of the metals and on the ionization state
of the enriched gas.  We will have more leverage to assess these
factors in a subsequent paper, in which we examine the role of
low-ionization metal absorption systems at $z \sim 6$ (Becker et al.,
in prep).  For now, however, we can offer some speculation based on
the \civ\ results alone.

Perhaps the most striking feature of the high-redshift \civ\ evolution
is how rapidly it proceeds at $z > 4.5$.  As noted above, both the
integrated comoving mass density (\Ociv) and the column density
distribution (\fN) remain nearly invariant over $2 \lesssim z \lesssim
4.5$.  Our redshift coverage extends down to $z = 5.3$, which means
that much of the buildup of \civ, at least for the weaker systems,
must happen over an interval $\Delta z \lesssim 1$, or $\Delta t
\lesssim 300$~Myr.  By comparison, neither the comoving star formation
rate density nor the stellar mass density is likely to increase by
more than a factor of three from $z \sim 5.3 - 6$ to $z \sim 4.5$
\citep{bouwens06,yan06,stark07,eyles07}.

\citet{oppenheimer06,oppenheimer08} have recently addressed the
observed \civ\ evolution using numerical simulations of large scale
structure, including various prescriptions for feedback from galactic
winds.  They find that a relatively constant \Ociv\ and \fN\ over $z
\sim 2-5$ can be produced by a steadily increasing IGM metallicity
offset with a declining \civ/C ratio due to energy input from winds.
Significantly, they find that in order to produce enough systems at $z
= 3.0-4.5$, the metals must largely be in place by $z = 4.5-6.0$.
Their best-fitting model \citep['vzw' from][]{oppenheimer06} uses
momentum-driven winds whose strength is tuned to strike the best
balance between metal enrichment and energy injection.  Convolving
their \fN\ from this model at $z \sim 4.5-6.0$ with our completeness,
and making a first order correction based on the fact that their
\Ociv\ declines by roughly a factor of two from $z \sim 4.5 - 5.3$ to
$z \sim 5.3-6.0$, we would expect to see $\sim 13$ \civ\ systems in
our data.  This is strongly excluded, which suggests that the mean
metallicity, the spatial distribution of metals, and/or the ionization
fraction of \civ\ at $z > 5.3$ must be evolving more rapidly than in
their simulations.

An important feature of the \citet{oppenheimer06} results is that
\civ\ traces increasingly smaller overdensities at higher redshifts
\citep[see also][]{simcoe06}.  The observed dropoff in \civ\ may
therefore indicate that most of the metals at $z > 5.3$ are confined
to overdensities greater than those where \civ\ is a sensitive tracer
of metallicity \citep[$-1 \lesssim \log{\rho/{\bar \rho}} \lesssim 1$
  at $z \sim 5.5$;][]{oppenheimer06}.  The metallicity gradient might
have to be fairly steep in order to suddenly recover the observed
number density of \civ\ systems at $z \sim 4.5$ by shifting the
\civ\ ``window'' to larger overdensities.  If the metal-enriched
regions are marginally self-shielded at $z > 5.3$, however, then a
moderate increase in the UV background may be sufficient to produce a
large increase in the number of \civ\ systems over a relatively small
$\Delta z$.  A hardening of the UV background could also produce an
increase in the \civ/C ratio.  \citet{madau08} have recently pointed
out that absorption in the Lyman series of \heii\ in the IGM can
substantially decrease the intensity of the UV background at energies
near the ionization potential of \ciii\ (3.5~Ryd).  If the volume
filling fraction of \heii\ is significantly smaller at $z \sim 4.5$
than at $z > 5.3$, as would be expected for an extended
\heii\ reionization process \citep[e.g.,][]{bolton08,mcquinn08}, then
the shape of the UV background may be more favorable to \civ\ at $z
\sim 4.5$ than at $z > 5.3$.

Given the strong \civ\ systems identified by \citet{simcoe06} and
\citet{r-w06}, it is especially surprising that we did not detect any
weak systems.  The two \civ\ systems towards SDSS~J1030$+$0524 are
marginally consistent with a power-law column density distribution
under the constraints provided by our data.  For example, for our 95\%
upper limit on the normalization of \fN\ for $\alpha = 1.8$
(Figure~\ref{fig:civ_plaw}), there is a 6\% chance that a single
sightline spanning $\Delta X = 3.1$ would contain at least two systems
with $13.8 < \log{N_{\mciv}} < 15.0$.  Among the six sightlines
included here and in the previous studies, however, there would be a
30\% chance that at least one sightline would contain two or more
strong systems.  This increases to 63\% for $\alpha = 1.0$, but in
either case the lack of detections along our four sightines would have
been highly unlucky.  It is intriguing, therefore, to consider the
possibility that there may be large-scale enrichment and/or ionization
fluctuations in the $z \sim 6$ IGM.

As noted by \citet{simcoe06}, there is good evidence that strong
\civ\ systems at lower redshifts are closely associated with actively
star-forming galaxies \citep{adelberger03,simcoe06b}.  If weak
\civ\ systems are absent from the $z \sim 6$ IGM because most of the
metal-enriched gas is in a lower ionization state, then the rare
strong \civ\ systems may point to regions near vigorously star-forming
galaxies, where the metallicity and the local ionizing background are
enhanced.  There is tentative evidence that this is the case from a
search for $z \sim 6$ Lyman break galaxies carried out by
\citet{kim08}.  They used HST/ACS to search for $i$-dropouts in the
fields of five $z \sim 6$ QSO.  Compared to the GOODS field, they find
overdensities in only two fields, with the strongest overdensity
towards SDSS~J1030$+$0524 \citep[first reported by][]{stiavelli05}.
The $i$-dropouts in that field appear spatially clustered, and their
$i-z$ colors are consistent with star-forming galaxies at $z \sim 5.8$
\citep{bouwens06}.  This suggests that the two strong \civ\ absorbers
seen towards SDSS~J1030$+$0524 may be associated with outflows from
UV-bright galaxies in a foreground filament.

In contrast, an underdensity of $i$-dropouts is seen towards
SDSS~J1148$+$5251, whose sightlines does not contain any $z > 5.3$
\civ\ absorbers.  This sightline is known to contain multiple $z \sim
6$ \oi\ systems \citep{becker06}.  The presence of metals in a
low-ionization state, combined with the lack of UV-bright galaxies,
suggests that this sightline was enriched at $z > 6$, but lacks a
sufficient amount of ongoing star formation to keep the enriched gas
highly ionized.  We will return to this point in a subsequent paper.
Finally, we note that the SDSS~J0836$+$0054 field contains an
overdensity of $i$-dropouts \citep{zheng06} yet does not show any
\civ\ or \oi\ \citep{becker06}.  If these galaxies are embedded in a
protocluster surrounding the radio-loud QSO, then hard UV photons from
the QSO itself may be over-ionizing any metal-enriched gas such that
it does not appear as \civ.  Spectroscopic follow-up of the
$i$-dropouts in QSO fields will help determine whether these scenarios
are correct.

\section{Conclusions}

We have presented a survey for \civ\ at $z = 5.3-6.0$ towards four $z
\sim 6$ QSOs using Keck/NIRSPEC in echelle mode.  This is the most
sensitive search for \civ\ at these redshifts to date, yet we find no
\civ\ systems in any of our spectra.  Based on our completeness
estimates and the column density distribution of \citet{songaila01},
which remains largely invariant over $2 < z < 4.5$, we would have
expected to detect roughly 13 systems in the range $13.0 <
\log{N_{\mciv}} < 14.5$ in the case of no evolution out to $z \sim 6$.
Our lack of detections implies that the number density of
\civ\ absorbers to which our data are sensitive declines by at least a
factor of $\approx 4.4$ (95\% confidence) from $z \sim 4.5$.

Our null result places strong constraints on the shape and
normalization of the \civ\ column density distributions at $ z = 5.3 -
6.0$.  For a power-law distribution, $f(N_{\mciv}) \propto
N_{\mciv}^{-\alpha}$, with a slope equal to that measured at $z \sim
3$ ($\alpha \approx 1.8$), the normalization must decline by at least
a factor of 4.4 (95\%).  The distribution may also flatten or steepen,
but only a narrow range of normalizations are allowed for each
$\alpha$ by our data and the fact that at least two strong
($\log{N_{\mciv}} > 13.8$) systems are seen in the two previously
studied sightlines \citep[][and confirmed herein]{simcoe06,r-w06}.
Our limits on \Ociv\ are less strong due to the fact that the comoving
density can be domiated by rare, strong systems.  Within the range
$13.0 < \log{N_{\mciv}} < 15.0$, however, the constraints on
\fN\ imply that \Ociv\ must decline by at least a factor of two from
$z \sim 2-4.5$ to $z = 5.3-6.0$ (see also Ryan-Weber et al., in prep).
The fact that \fN\ and \Ociv\ are relatively constant at $z \lesssim
4.5$ further means that most of the buildup of \civ\ in the IGM must
occur over an interval $\Delta z \lesssim 1$, or $\Delta t \lesssim
300$~Myr.

The fact that \civ\ in the IGM rapidly increases between $z \sim 5.3$
and $z \sim 4.5$ suggests that ionization effects may be playing an
important role.  That is, a significant fraction of intergalactic
metals we see as \civ\ at $z \sim 4.5$ may already be in place at $z >
5.3$, but in a lower ionization state.  This may especially be true if
the metals are confined to relatively high overdensities
\citep[e.g.,][]{oppenheimer06}, but other factors may also be at work.
An increase in the overall IGM metallicity, a wider dispersal of
metals by galactic winds, and an increase in the \civ/C ratio due to
an increase in the intensity or a hardening of the UV background would
all help to explain the rapid increase in \civ\ at $z < 5.3$.  To
evaluate the relative importance of these effects, it is critical to
measure the abundance of metals in lower ionization states.  We will
do this in a subsequent work.

The contrast between our lack of detections along four sightlines and
the presence of two strong \civ\ absorbers towards SDSS~J1030$+$0524
suggests that there may be large-scale variations in IGM enrichment
and/or in the ionization state of the metal-enriched gas.  We are
currently planning follow-up observations of SDSS~J1030$+$0524 to look
for weak \civ\ systems.  The presence of additional \civ\ would more
strongly indicate that this sightline is either more enriched or more
highly ionized than those in our current sample.  Alternatively, the
strong \civ\ systems may be associated with rare, UV-bright
star-forming galaxies, in which case the enhancement of \civ\ would be
a local effect.  This is supported by the overabundance of
$i$-dropouts in the field of SDSS~J1030$+$0524
\citep{stiavelli05,kim08}, although spectroscopic follow-up is
required to be certain that these galaxies are associate with the
\civ\ systems.

Finally, we note that the decline of \civ\ at $z > 5.3$ poses an
interesting challenge to the use of low-ionization metal lines as
tracers of reionization.  If the number density of \civ\ systems had
remained constant out to $z \sim 6$, then we would be more certain
that the IGM is sufficiently enriched at early times for metal lines
to serve as meaningful probes of the ionization state of the gas.  The
disappearance of \civ\ at high redshifts, however, means that if an
``\oi\ forest'' is not found at $z \gtrsim 6$, the result could be
attributed to a deficit of metals rather than to a highly-ionized IGM.
On the other hand, the detection of a significant number of neutral
metal absorption systems at $z \sim 6$ would have a double
significance: that the IGM is at least partially metal-enriched by
that redshift, and that pockets of the IGM were not permanently
ionized until well after the early reionization epoch favored by CMB
studies.  This will be the focus of the following paper.


\vspace*{5.8mm}   

The authors would like to thank Emma Ryan-Weber and Max Pettini for
many stimulating conversations over the course of this work; as well
as Martin Haehnelt, Bob Carswell, and Rob Simcoe for their comments on
the first draft of this paper.  G.~B. was supported by the Kavli
Institute for Cosmology at the Institute of Astronomy in Cambridge.
M.~R. was supported by the National Science Foundation through grant
AST 05-06845.

\end{document}